\documentclass[final]{svjour2}
\usepackage{graphicx}
\usepackage{rotating}
\usepackage{amssymb}
\usepackage{mathptmx}
\usepackage[numbers]{natbib}
\makeatletter
\journalname{Journal of Low Temperature Physics}

\bibpunct{}{}{,}{s}{}{,}

\newcommand{\beq}{\begin{equation}}
\newcommand{\eeq}{\end{equation}} 
\newcommand{\beqa}{\begin{eqnarray}}
\newcommand{\eeqa}{\end{eqnarray}}
\newcommand{\ba}{\begin{array}}
\newcommand{\ea}{\end{array}}

\begin{document}

\newcommand{\hdblarrow}{H\makebox[0.9ex][l]{$\downdownarrows$}-}
\title{Dispersive effects in the unitary Fermi gas}
\author{ F. Ancilotto$^{1,2}$, L. Salasnich$^{1}$ and F. Toigo$^{1,2}$} 
\institute{$^1$Dipartimento di Fisica e 
Astronomia ``Galileo Galilei'' and CNISM, 
Universit\`a di Padova, Via Marzolo 8, 35122 Padova, Italy \\
$^2$CNR-IOM Democritos, via Bonomea, 265 - 34136 Trieste, Italy\\
\email{ francesco.ancilotto@unipd.it, luca.salasnich@unipd.it,
flavio.toigo@unipd.it}}

\date{\today}

\maketitle

\begin{abstract} 
We investigate within density functional
theory various physical properties of the zero-temperature 
unitary Fermi gas  
which critically depend on the presence of a 
dispersive gradient term in the equation of state. 
First, we consider the unitary 
Fermi superfluid gas confined to a semi-infinite domain and calculate 
analytically its density profile and surface tension. 
Then we study the quadrupole modes of the superfluid 
system under harmonic confinement finding a reliable 
analytical formula for the oscillation frequency, which reduces 
to the familiar Thomas-Fermi one in the limit of a large number 
of atoms. Finally, we discuss the formation and propagation 
of dispersive shock waves in the collision between two resonant 
fermionic clouds, and compare our findings with recent
experimental results.
\end{abstract} 

\section{Introduction}

In the last years the crossover from the weakly paired 
Bardeen-Cooper-Schrieffer 
(BCS) state to the Bose-Einstein condensate (BEC) 
of molecular dimers with ultra-cold two-hyperfine-components Fermi
vapors atoms has been investigated 
by several experimental and theoretical groups \cite{zwerger}.
When the densities of the two spin components are equal, and when the 
gas is dilute so that the range of the inter-atomic potential
is much smaller than the inter-particle distance, then the interaction 
effects are described by only one parameter: the s-wave scattering 
length, whose sign determines the 
character of the gas. Fano-Feshbach 
resonances can be used to change the value and the sign of the 
scattering length, simply by tuning an external magnetic field. 
At resonance the scattering length diverges so that the gas 
displays a very peculiar character, being at the same 
time dilute and strongly interacting. In this regime all scales 
associated with interactions disappear from the problem 
and the energy of the system is expected to be proportional to that 
of a non interacting fermions system. This 
is called the unitary regime \cite{zwerger,bulgac1}. 

Recently it has been remarked \cite{bulgac1} that 
the superfluid unitary Fermi gas, 
characterized by a divergent s-wave scattering length \cite{zwerger}, 
can be efficiently described at zero temperature by 
phenomenological density functional
theory. Indeed, different theoretical groups have proposed various 
density functionals. For example Bulgac and Yu have introduced a superfluid 
density functional based on a Bogoliubov-de Gennes 
approach to superfluid fermions 
\cite{bulgac2,bulgac3}. Papenbrock and Bhattacharyya \cite{papenbrock} 
have instead proposed a Kohn-Sham density 
functional with an effective mass to take into account nonlocality effects. 
Here we adopt instead the extended Thomas-Fermi functional of the 
unitary Fermi gas that we have proposed few years ago \cite{salasnich}.
The total energy in the extended Thomas-Fermi functional 
contains a term proportional to the kinetic
energy of a uniform non interacting gas of fermions
with number density $n({\bf r})$, plus 
a gradient correction of the form $\lambda \hbar^2/(8m) 
(\nabla n/n)^2$, originally introduced by
von Weizs\"acker to treat surface effects in nuclei \cite{von}, 
and then extensively applied to study electrons \cite{vari-uffa}, 
showing good agreement with Kohn-Sham calculations. 
In the context of the BCS-BEC crossover, the presence of
the gradient term (and its actual weight) 
is a debated issue \cite{v1,v2,v3,v4,v5,v6,v7,v8}. 
The main advantage of using such a functional is that, as it
depends only on a single function of the coordinates, i.e. the
order parameter, can be used to study systems with quite large 
number of particles $N$. Other functionals, which are based instead on 
single-particle orbitals, require 
self-consistent calculations with a numerical load 
rapidly increasing with $N$. 

In the last years we have successfully applied our extended Thomas-Fermi 
density functional and its time-dependent version 
\cite{salasnich} to investigate 
density profiles \cite{salasnich,anci1}, 
collective excitations \cite{anci1}, Josephson effect 
\cite{anci2} and shock waves 
\cite{salasnich-shock,anci3} of the unitary Fermi gas. In addition, 
the collective modes of our density functional have been used to study 
the low-temperature thermodynamics of the unitary Fermi gas 
(superfluid fraction, first sound and second sound) \cite{salasnich-thermo}
and also the viscosity-entropy ratio of the unitary Fermi gas 
from zero-temperature elementary excitations \cite{salasnich-ratio}. 

For the superfluid unitary Fermi gas one expects 
the coexistence of dispersive and dissipative terms 
in the equation of state \cite{anci3,salasnich-thermo,salasnich-ratio,manas}. 
However, at zero temperature only dispersive term 
survive \cite{anci3,landau}. In this paper we investigate various physical 
quantities of the unitary Fermi gas, which depend on the presence of a 
dispersive gradient term in the zero-temperature equation of state. 
In the fist part we investigate static properties like density profiles 
and surface tension. In the second part we study dynamical properties 
like collective oscillations and dispersive shock waves due 
to the collision between two fermionic clouds. 

\section{Extended Thomas-Fermi density functional} 

The energy density of a uniform Fermi gas at unitarity 
depends on the constant density $n$ as follows \cite{zwerger} 
\beq
\label{eq:energy1}
{\cal E}_{unif}(n) = \xi \frac{3}{5} \frac{\hbar^2}{2m}(3\pi^2)^{2/3} 
\, n^{5/3} \,  
\eeq
where $\xi \simeq 0.4$ is a universal parameter of 
the Fermi gas \cite{zwerger}. 
In the presence of an external trapping potential $U({\bf r})$ 
the Hohenberg-Kohn theorem \cite{hkt} ensures that 
the ground-state local density $n({\bf r})$ of the system 
can be obtained by minimizing the energy functional 
\beq
\label{eq:ETF} 
E[n] = F[n] + \int  U({\bf{r}}) \ n({\bf r}) \ d^3{\bf{r}} \, 
\eeq
where $F[n]$ is a (generally unknown) energy functional of the 
internal energy, which is independent on $U({\bf r})$. 
In our extended Thomas-Fermi approach we choose 
\beq
\label{eq:energy2}
F[n] =  \int {\cal E}(n,\nabla n) \ d^3{\bf{r}} \,  
\eeq
where the local energy density ${\cal E}(n,\nabla n)$ is given by 
\beq 
{\cal E}(n,{\nabla} n) = {\cal E}_{unif}(n) + 
\lambda {\hbar^2 \over 8 m} {(\nabla n)^2\over n}    
\label{ee-dft}
\eeq
In this expression, which is the equation of state (internal energy density) 
of the inhomogeneous system, the first term is the Thomas-Fermi-like term 
describing the uniform system, while 
the second term is the von Weizs\"acker gradient correction \cite{von}. 
We have obtained the value $ \lambda \simeq 0.25 $ 
by fitting accurate Monte Carlo results
for the energy of fermions confined in
a spherical harmonic trap close to unitary 
conditions \cite{salasnich,miomao}.

\section{Semi-infinite domain: density profile and surface tension}

We put in evidence the effects of 
the gradient term of our density functional 
by considering the external potential 
\beq 
U({\bf r}) = \left\{ 
\ba{ccc} 
+ \infty & \mbox{ for } & z < 0 \\
0 & \mbox{for}   & z> 0 \\
\ea
\right. 
\label{potential}
\eeq
acting on the Fermi superfluid. This implies that the superfluid density 
$n({\bf r})$ must go to zero at the boundary $z=0$, while it becomes a 
constant $\bar{n}= \frac{1}{3 \pi ^2} 
(\frac{2 m \bar{\mu}}{ \xi  \hbar ^2})^{\frac{3}{2}}$ far from the boundary. 
Since in the present calculations we 
fix the chemical potential $\bar{\mu}$ rather than the 
total number of fermions $N$, it is useful to introduce 
the zero-temperature grand potential energy functional 
of the unitary Fermi gas  
\beqa 
\Omega &=& \int \left[ {\cal E}(n,\nabla n) + U({\bf r}) n({\bf r}) 
- \bar{\mu} \, n({\bf r})  \right] \ d^3{\bf r}    
\label{e-dft}
\eeqa 

It is convenient to introduce the characteristic length of the system
\beq 
l_s = {\sqrt{\lambda/\xi} \over (3\pi^2 \bar{n})^{1/3}} 
\label{length}
\eeq
and to rewrite the density in terms of a function of the 
adimensional variable $\zeta = {z/l_s}$ as 
\beq 
\sqrt{n({\bf r})} = \sqrt{\bar{n}} \ f(\frac{z}{l_s}) \ = 
\sqrt{\bar{n}} \ f(\zeta)  
\label{density}
\eeq
The grand-potential (\ref{e-dft}) then becomes:
\beq
 \Omega = A \, l_s \,{\bar n} \, { \bar \mu}\int_0^{L/l_s} 
\Big[ f'(\zeta)^2 + {3\over 5} f(\zeta)^{10/3} 
- f(\zeta)^2 \Big] \ d\zeta  
\label{omega}
\eeq
where $f'(\zeta) = df/d\zeta$. $A$ is the area in the $(x,y)$ 
plane and $L$ is the length in the $z$ direction. 
The function $f(\zeta)$ minimizing the 
grand-potential obeys the equation:
\beq 
- f''(\zeta ) + f(\zeta )^{7/3} = f(\zeta )  
\label{super}
\eeq
with the boundary conditions 
$
f(0) = f'(+\infty) = 0 \quad \mbox{ and } \quad f(+\infty) = 1 \; . 
$
The first integral of the system is  
\beq 
K = {1\over 2} f'(\zeta)^2 + {1\over 2} f(\zeta )^2 
- {3\over 10} f(\zeta )^{10/3}  
\label{constant0}
\eeq
By using the appropriate boundary conditions we find 
$K = {1\over 5}$ and $f'(0) = \sqrt{2\over 5}$.
It is then straightforward to get the integral equation  
\beq 
\zeta = \int_0^{f(\zeta)} 
{df \over \sqrt{{2\over 5}+{3\over 5}f^{10/3}-f^2} }  
\label{f-exact}
\eeq
which gives implicitly the profile function $f(\zeta)$. 

\begin{figure}
\begin{center}
\includegraphics[width=0.7\linewidth]{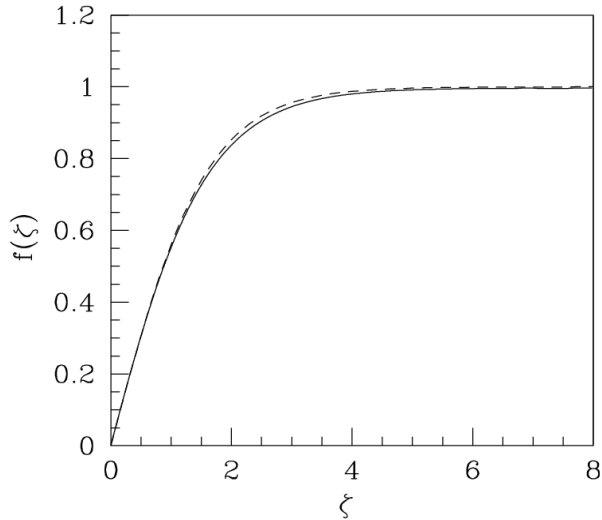} 
\end{center}
\caption{Scaled density profile $f(\zeta)$ 
of the unitary Fermi gas in the semi-infinite domain. 
Solid line: numerical integration of Eq. (\ref{f-exact}). 
Dot-dashed line: analytical interpolation
$f(\zeta) =\tanh{(\sqrt \frac{2}{5} \zeta)}$.
} 
\label{fig1}
\end{figure}

In Fig. \ref{fig1} we plot the scaled density profile $f(\zeta)$ 
obtained from the numerical integration (solid line) 
of Eq. (\ref{f-exact}), together with 
the function $f(\zeta) =\tanh{(\sqrt \frac{2}{5} \zeta)}$ (dot-dashed line), 
which provides an excellent overall approximation to it.

In the limit $L/l_s\to +\infty$ the grand potential energy $\Omega/A$
is divergent due to the asymptotic energy 
\beq 
{\Omega_{asy} \over A} = -{2\over 5}  
\sqrt{\lambda \xi} \,  \bar{n}^{4/3}
(3\pi^2)^{1/3}{\hbar^2\over 2m} \int_0^{L/l_s}d\zeta 
\eeq 
where $f(\zeta) = 1$ and $f'(\zeta)=0$. 
The surface tension $\sigma$ is defined as 
\beq 
\sigma = {(\Omega - \Omega_{asy})\over A} 
\eeq 
in the limit $L/l_s\to +\infty$, i.e. 
\beq 
\sigma = \sqrt{\lambda \xi} \,  \bar{n}^{4/3} 
(3\pi^2)^{1/3}{\hbar^2\over 2m} 
\int_0^{+\infty} 
\Big[ f'(\zeta)^2 + {3\over 5} f(\zeta)^{10/3} 
- f(\zeta)^2 + {2\over 5} \Big] \ d\zeta  
\eeq
Eq. (\ref{constant0}) with $K=1/5$, and  
$d\zeta = df/f'(\zeta)$, helps us to rewrite the 
formula of the surface tension as 
\beq 
\sigma =  I \  \sqrt{\lambda \xi} \, \bar{n}^{4/3} 
(3\pi^2)^{1/3}{\hbar^2\over 2m}   
\eeq
where $I = 2 \int_0^1 \sqrt{{2\over 5} + {3\over 5}f^{10/3} - f^2} \ df =0.82$. 

\begin{figure}
\begin{center}
\includegraphics[width=0.7\linewidth]{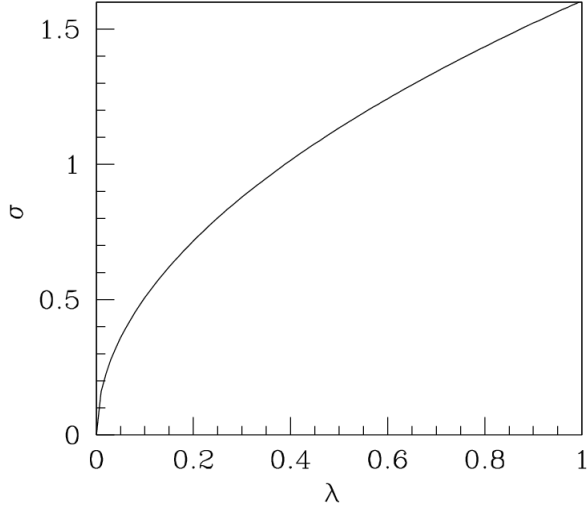} 
\end{center}
\caption{Surface tension $\sigma$ as a function 
of the von Weizs\"acker gradient coefficient $\lambda$. 
$\sigma$ is in units of $\bar{n}^{4/3} \hbar^2/(2m)$.} 
\label{fig2}
\end{figure}

In Fig. \ref{fig2} we plot the surface tension $\sigma$ as a function 
of the von Weizs\"acker gradient coefficient $\lambda$ setting 
$\xi =0.4$ \cite{zwerger,salasnich}. 
By using $\lambda=1/4$, which is a reasonable estimate 
\cite{salasnich,best} consistent with Monte 
Carlo results \cite{salasnich,anci2}, 
one finds 
$
\sigma = 0.80\, \bar{n}^{4/3} {\hbar^2\over 2 m}  
$ 
This value is of the same order of magnitude 
of previous microscopic determinations of the surface tension 
$\sigma$ based on different theoretical approaches\cite{caldas,mueller,andrei}
and different boundary conditions at the interface. 

\section{Extended superfluid hydrodynamics}

Let us now consider dynamical properties of the unitary Fermi gas. 
The starting point are the zero-temperature equations of superfluid 
hydrodynamics \cite{landau}, given by 
\beqa
{\partial n \over \partial t} + { 
\nabla} \cdot (n {\bf v}) = 0  
\label{hy1}
\\
m{\partial {\bf v} \over \partial t} + { 
\nabla} [ {m\over 2} v^2 + 
{\partial {\cal E}\over \partial n}- { \nabla}\cdot 
{\partial {\cal E}\over \partial ({ \nabla} n)}+U({\bf r})] = {\bf 0}   
\label{hy2}
\eeqa
where where $n({\bf r},t)$ is the time-dependent scalar density field 
and ${\bf v}({\bf r},t)$ is the time-dependent vector velocity field. 
In the case of the unitary Fermi gas the
local energy density ${\cal E}$ can be written by using Eq. (\ref{ee-dft}). 
If $\lambda =0$, then Eqs. (\ref{hy1}) and (\ref{hy2}) reproduce 
by construction the familiar 
equations of superfluid hydrodynamics \cite{zwerger}. 

For a fermionic superfluid 
the velocity is irrotational, i.e. 
${\bf \nabla}\wedge {\bf v}={\bf 0}$, and the circulation 
is quantized, i.e. 
\beq 
\oint {\bf v} \cdot d{\bf r} = {\hbar \over m}\, \pi \, l   
\eeq
where $l$ is an integer quantum number. 
This means that the velocity field of the unitary Fermi gas 
can be written as 
\beq 
{\bf v}({\bf r},t) = {\hbar \over 2m}\nabla \theta({\bf r},t)  
\eeq
where $\theta({\bf r},t)$ is the phase 
of the Cooper pairs \cite{zwerger, best}. 
Eqs. (\ref{hy1}) and (\ref{hy2}) can be interpreted 
as the Euler-Lagrange equations of the following action functional 
\beq 
A = \int  \left[ 
{\hbar\over 2} {\dot{\theta}} \, n 
+ {\hbar^2\over 8m} (\nabla \theta)^2 \, n 
+ {\cal E}(n,{\nabla} n) + U({\bf r}) \, n \right] \ d^3{\bf r}\ dt    
\label{e-popov}
\eeq
which depends on the local density $n({\bf r},t)$ and the phase 
$\theta({\bf r},t)$. In the next section we use 
this action functional (\ref{e-popov}) to 
investigate zero-temperature collective oscillations 
of the unitary Fermi gas trapped by an external potential. 

\section{Quadrupole oscillations}

\begin{figure}
\begin{center}
\includegraphics[width=0.8\linewidth]{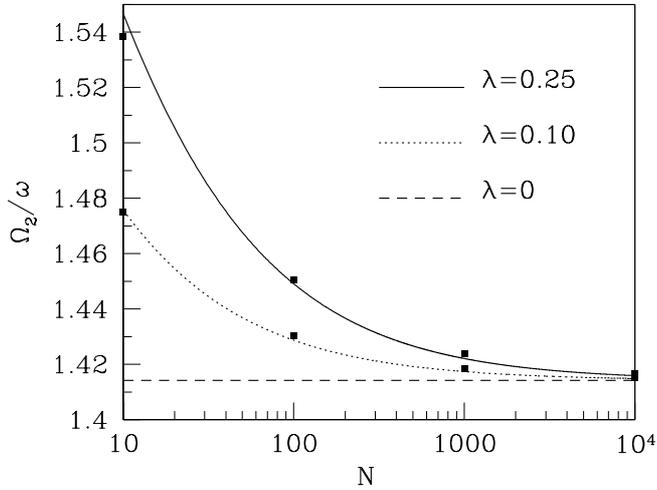} 
\end{center}
\caption{Quadrupole frequency $\Omega_2$ of 
the unitary Fermi gas with $N$ atoms under harmonic 
confinement of frequency $\omega$. Three different 
values of the gradient coefficient ${\lambda}$. 
For $\lambda=0$ (TF limit): $\Omega_2=\sqrt{2} \omega$. 
Filled squares are the numerical results, while the 
curves are obtained with the analytical formula, Eq. (\ref{very-good}). }
\label{fig3}
\end{figure}

In the case of spherically-symmetric harmonic confinement, i.e.   
\beq
U({\bf r})= {1\over 2} m \omega^2 r^2 \; , 
\eeq
we have numerically studied \cite{anci1} 
the collective modes of the unitary Fermi gas 
for different number $N$ of atoms by means of 
Eqs. (\ref{hy1}) and (\ref{hy2}). 
As predicted by Y. Castin \cite{castin}, the frequency $\Omega_0$ of the 
monopole mode (breathing mode) 
and the frequency $\Omega_1$ dipole mode (center of mass 
oscillation) do not depend on $N$: 
\beq 
\Omega_0 = 2 \omega  \quad\quad \mbox{ and } \quad\quad \Omega_1 
= \omega    
\eeq
We have found \cite{anci1} instead that the frequency 
$\Omega_2$ of the quadrupole ($l=2$) mode 
depends on $N$ and on the choice of the gradient 
coefficient $\lambda$. In particular, in Ref. \cite{anci1} we 
have calculated $\Omega_2$ for increasing values of $N$ up to 
$N=100$. 

In this paper we extend our previous calculations by taking 
into account much larger values of $N$. To excite the 
quadrupole mode, we solve numerically 
Eqs. (\ref{hy1}) and (\ref{hy2}) with the initial condition 
\beqa 
n({\bf r},t=0) &=& n_{gs}({\bf r}) \, , 
\\
{\bf v}({\bf r},t=0) &=& {\hbar\over 2m} {\bf \nabla}
\Big[ \epsilon (2z^2 - x^2 - y^2) \Big] \; ,   
\eeqa
where $n_{gs}({\bf r})$ is the ground-state 
density profile and $\epsilon$ a small parameter. 
The results are shown in Fig. \ref{fig3} where we plot 
the quadrupole frequency $\Omega_2$ as a function of the number 
$N$ of atoms for three values of the gradient coefficient $\lambda$. 

The numerical results of the figure can 
quite well be captured by a time-dependent 
Gaussian variational approach \cite{sala-tdva}. We set 
\beq 
n(x,y,z,t) = {N\over \pi^{3/2}\sigma_{\bot}(t)^2\sigma_z(t)}\, 
e^{-(x^2+y^2)/\sigma_{\bot}(t)^2} \ e^{-z^2/\sigma_z(t)^2}  
\eeq
and
\beq
\theta(x,y,z,t) = \beta_{\bot}(t) (x^2+y^2) + \beta_z(t) z^2   
\eeq
where $\sigma_{\bot}(t)$, $\sigma_z(t)$, $\beta_{\bot}(t)$, 
and $\beta_z(t)$ are the time-dependent variational parameters. 
Inserting these quantities into the action functional 
(\ref{e-popov}), after integration over space variables we find  
\beq 
A = N \int  L(\sigma_{\bot}(t), \sigma_z(t),
\beta_{\bot}(t),\beta_z(t)) \ dt  
\eeq
where the effective Lagrangian $L$ is given by  
$$
L= 
{\hbar\over 2} 
\left( {\dot \beta_{\bot}}\sigma_{\bot}^2 + {1\over 2}
{\dot \beta_z}\sigma_z^2 \right) 
+ 
{\hbar^2\over 2m}
\left( \sigma_{\bot}^2 \beta_{\bot}^2 + {1\over 2}
\sigma_{z}^2 \beta_{z}^2 \right)
$$
\beq 
+ 
{\lambda \hbar^2\over 2m}
\left(  {1\over \sigma_{\bot}^2}+ {1\over 2\sigma_z^2} \right) 
+ {1\over 2} 
m \omega^2 
\left( \sigma_{\bot}^2 + {1\over 2} \sigma_z^2 \right) 
+ 
{\hbar^2\over 2m} N^{2/3} {g\over (\sigma_{\bot}^2\sigma_z)^{2/3}} 
\eeq
with $g=(3/5)^{5/2}(3\pi^2)^{2/3}\xi/\pi$.
From this effective Lagrangian we calculate analytically the 
small oscillations around the equilibrium configuration, following 
the procedure described in Ref.\cite{sala-tdva}. 
In this way we get the following formula for the 
frequency of the quadrupole mode: 
\beq
\Omega_2 = \omega \ \sqrt{2 + 6\, {\lambda \over g N^{2/3}}
\over 1 + {3\over 2} {\lambda \over g N^{2/3}}} \; ,
\label{very-good}
\eeq
In the limit $N\to \infty$ it gives the Thomas-Fermi result \cite{zwerger} 
$\Omega = \sqrt{2}\omega$, 
while in the limit $N\to 0$ it gives 
$\Omega = 2 \omega$, which is the quadrupole oscillation frequency 
of non-interacting atoms \cite{zwerger}. The curves reported 
in Fig. \ref{fig3} show that this analytical formula 
reproduces remarkably well the numerical results (filled squares). 
This fact suggests the possibility of using 
Eq. (\ref{very-good}) to determine the value of $\lambda$ 
from experimental measurements of $\Omega_2$.

\section{Collision of resonant Fermi clouds}

\begin{figure}
\begin{center}
\includegraphics[width=0.9\linewidth]{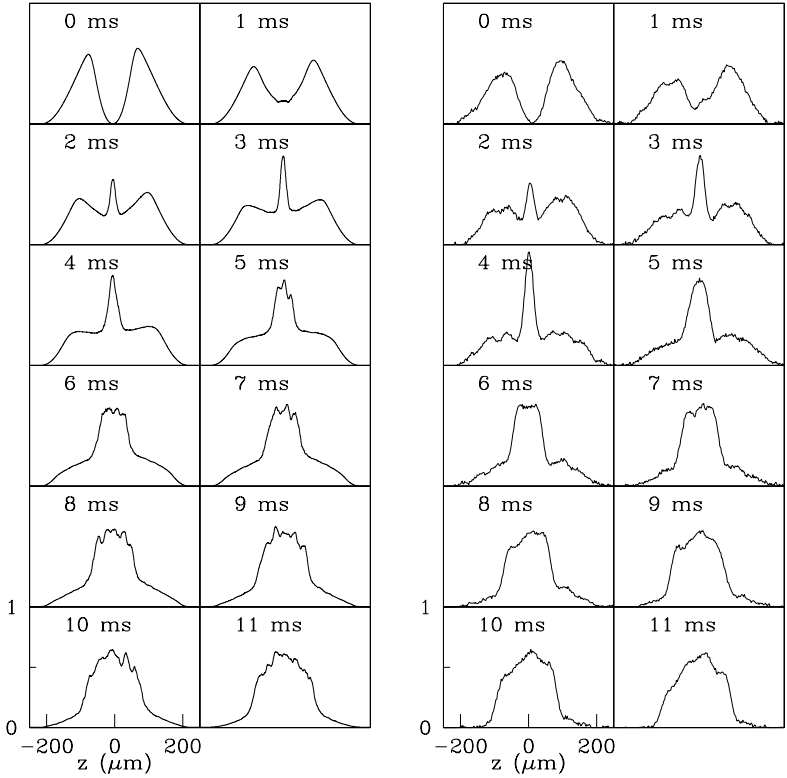}
\end{center}
\caption{1D density profiles at different times $t$ 
showing the collision of two strongly interacting
Fermi clouds. Left part: our calculations \cite{anci3}.
Right part: experimental data from Ref. \cite{thomas}. 
The normalized density is in units of $10^{-2}/\mu m$ per particle.}
\label{fig4}
\end{figure}

We use here the generalized superfluid hydrodynamics equations
to obtain the long-time dynamics of the collision between two 
initially separated Fermi clouds, simulating  
the experiments of Ref. \cite{thomas}. The details of our 
calculations can be found in 
Ref. \cite{anci3}. Here we summarize the main results 
of our study. In our simulations we tried to reproduce as closely as 
possible the experimental conditions of Ref. \cite{thomas} 
which we summarize briefly in the following. 
A 50:50 mixture of the two lowest hyperfine states of $^6$Li
(for a total of $N=2\times 10^5$ atoms)
is confined by an axially symmetric cigar-shaped laser trap,
elongated along the z-axis. The resulting trapping potential is  
$U(r,z)=0.5m[\omega _r^2 r^2+ \omega _z^2z^2]$, where 
$r^2=x^2+y^2$,
$\omega _r=2\pi \times 437$ Hz and $\omega _z=2\pi \times 27.7$ Hz.
The trapped Fermi cloud is initially 
bisected by a blue-detuned beam which provides a
repulsive knife-shaped potential.
This potential is then suddenly removed, allowing for the
two separated parts of the cloud to collide with each other. 
The system is then let to evolve for a given hold time
$t$ after which the trap in the radial direction  is removed.
The system is allowed to evolve for 
another $1.5$ ms during which the gas expands in the r-direction
(during this extra expansion time, the confining trap frequency along
the z-axis is changed to $\omega _z=2\pi \times 20.4$ Hz),
and finally a (destructive) image of the cloud 
is taken. The process is repeated from the beginning for 
another different value for the hold time $t$. 
The main effect observed in the experiment \cite{thomas} is the 
presence of shock waves, i.e. of regions characterized by
large density gradients, in the colliding clouds.
The experimental results are shown in the
right part of Fig. \ref{fig4} as a sequence of
one-dimensional profiles obtained by averaging
along one transverse direction the observed cloud density
at different times.

\begin{figure}
\begin{center}
\includegraphics[width=0.7\linewidth]{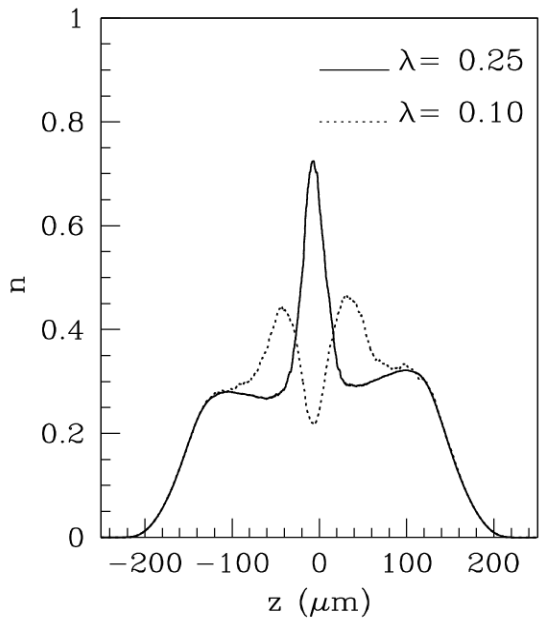} 
\end{center}
\caption{Calculated density profiles after 3 ms.
The solid line shows the results obtained using $\lambda =0.25$,
i.e. the same value used to compute the 
sequence shown in the left part of Fig.\ref{fig4}.
The dotted line shows instead the result obtained using $\lambda =0.1$.}
\label{fig5}
\end{figure}

We simulated the whole procedure by using 
the Runge-Kutta-Gill fourth-order method \cite{rkg,pi} 
to propagate in time the solutions of the following 
non-linear Schr\"odinger equation (NLSE) 
\beq 
i \hbar {\partial \over \partial t} \Psi = 
\Big[ -{\hbar^2 \over 4 m} \nabla^2 + 2 U({\bf r}) + 
2 {\hbar^2\over 2m}(3\pi^2)^{2/3} \xi |\Psi|^{4/3} 
+(1 - 4 \lambda ){\hbar^2\over 4m} 
{\nabla^2 |\Psi|\over |\Psi|} \Big] \Psi 
\label{nlse}
\eeq 
which is strictly equivalent \cite{salasnich,best} 
to Eqs. (\ref{hy1}) and (\ref{hy2}),  
with ${\cal E}(n,\nabla n)$ given by Eq. (\ref{ee-dft}), and
\beq 
\Psi({\bf r},t) = \sqrt{n({\bf r},t)}\ e^{i\theta({\bf r},t)}   
\eeq
Since the confining potential used in the experiments
is cigar-shaped, we have exploited the resulting cylindrical symmetry 
of the system by representing the solution of our NLSE on a 2-dimensional
$(r,z)$ grid. During the time evolution of our system, 
when the two clouds start to overlap, many ripples 
whose wavelength is comparable to the interparticle distance
are produced in the region of overlapping densities. 
In order to properly compare our results with the 
experimental data of resonant fermions \cite{thomas}, 
which are characterized by a finite
spatial resolution, we smooth the calculated profiles at 
each time $t$ by local averaging the density within a space window 
of $\pm 5$ $\mu m$ centered around the calculated point.

The results of our simulations, for the whole
time duration of the experiments, and after 
the smoothing procedure is applied to the (y-averaged) density 
profile at each time, are shown 
in the left part of Fig. \ref{fig4} (see also Ref.\cite{anci3}),
plotted along the long trap axis, for the same
time frames as in the experiment. 
Remarkably, there is a striking correspondence between the experimental data 
and the results of our simulation. 
At variance with the current interpretation of
the experiments, where the role of viscosity is
emphasized\cite{thomas}, we obtained a quantitative agreement
with the experimental observation of the dynamics of the
cloud collisions within our superfluid
effective hydrodynamics approach, where density
variations during the collision are controlled by a purely dispersive
quantum gradient term. 

We find that changing $\lambda$ from the
optimal value $\lambda =1/4$ has profound consequencies on the
long-time evolution of the colliding clouds,
providing density profiles which are completely
different from the experimental ones.
This is shown in Fig.\ref{fig5}, where the 
simulated density profile after 3 ms is 
calculated using two different values for $\lambda $. 
Such a strong dependence on $\lambda$ of the time evolution of
a Fermi cloud made of a large number of atoms
is at first sight surprising, because 
the gradient term should become less and less
important with increasing $N$.
We believe that such dependence is due to the
presence of shock waves (i.e. regions characterized by
large density gradients) in the colliding clouds,
as discussed in Ref.\cite{anci3}.

\section{Conclusions}

In the first part of the paper we have calculated the density profile 
and surface tension $\sigma$ of the superfluid unitary 
Fermi gas in a semi-infinite domain 
by using the extended Thomas-Fermi density functional,  
where the surface effects are modelled by the the von Weizs\"acker 
gradient term. Indeed we have found that $\sigma$ is proportional 
to $\sqrt{\lambda}$, where $\lambda$ is the phenomenological coefficient 
of the von Weizs\"acker term. 
In the second part of the paper we have investigated 
dispersive dynamical effect which crucially depend on the presence of 
a gradient term in the equation of state of the unitary Fermi gas. 
In particular, we have studied the quadrupole modes of the superfluid 
system under harmonic confinement, finding a reliable 
analytical formula for the oscillation frequency, which reduces 
to the familiar Thomas-Fermi one in the limit of a large number 
of atoms. Finally, we have numerically studied the
long-time dynamics of shock waves in the ultracold unitary Fermi gas. 
Two main results emerge from our calculations:
a) at zero temperature the simplest regularization 
process of the shock is purely dispersive, mediated by the 
quantum gradient term, 
which is one of the ingredient in our DF approach;
b) the quantum gradient term plays an important role 
not only in determining the static density profile of 
small systems, where surface effects are important, 
but also in the fast dynamics of large systems, 
where large density gradients may arise. 

\begin{acknowledgements}
We thank James Joseph, John E. Thomas and 
Manas Kulkarni for fruitful discussions. 
\end{acknowledgements}

\end{document}